\documentclass[a4paper,11pt,oneside]{elsarticle}
\usepackage{amsmath}
\usepackage{graphicx}
\usepackage{epstopdf}
\usepackage{amsmath}
\usepackage{amsfonts}
\usepackage{amsthm}
\usepackage{mathrsfs}
\usepackage{subcaption}
\usepackage{amssymb}
\usepackage{physics}
\usepackage{xcolor}
\usepackage{caption}
\usepackage{subcaption}

\makeatletter
\def\ps@pprintTitle{%
 \let\@oddhead\@empty
 \let\@evenhead\@empty
 \def\@oddfoot{}%
 \let\@evenfoot\@oddfoot}
\makeatother

\newcommand{\nab}{\nabla^2}

\newcommand{\p}{\partial}

\newcommand{\De}{D_\mathrm{e}}
\newcommand{\Ds}{D_\mathrm{s}}
\newcommand{\cs}{c_\mathrm{s}}
\newcommand{\ce}{c_{\mathrm{e}}}

\newcommand{\bce}{\bar{c}_\mathrm{e}}
\newcommand{\bcen}{\bar{c}_{\mathrm{e,n}}}
\newcommand{\bcep}{\bar{c}_{\mathrm{e,p}}}

\newcommand{\bPhis}{\bar{\Phi}_\mathrm{s}}
\newcommand{\bPhie}{\bar{\Phi}_\mathrm{e}}
\newcommand{\Phis}{\Phi_\mathrm{s}}
\newcommand{\Phie}{\Phi_\mathrm{e}}

\newcommand{\bPhisp}{\bar{\Phi}_\mathrm{s,p}}
\newcommand{\bPhisn}{\bar{\Phi}_\mathrm{s,n}}
\newcommand{\bPhiep}{\bar{\Phi}_\mathrm{e,p}}
\newcommand{\bPhien}{\bar{\Phi}_\mathrm{e,n}}
\newcommand{\Omegap}{\Omega_\mathrm{p}}
\newcommand{\Omegan}{\Omega_\mathrm{n}}
\newcommand{\OmegaS}{\Omega_\mathrm{s}}
\newcommand{\Omcell}{\Omega_\mathrm{cell}}
\newcommand{\Ombatt}{\Omega_\mathrm{batt}}
\newcommand{\OCP}{U_\mathrm{OCP}}
\newcommand{\arsinh}{\mathrm{arsinh}}
\newcommand{\alphas}{\alpha_\mathrm{s}}
\newcommand{\alphae}{\alpha_\mathrm{e}}
\newcommand{\mrs}{\mathrm{s}}
\newcommand{\mre}{\mathrm{e}}
\newcommand{\Ln}{L_\mathrm{n}}
\newcommand{\Lp}{L_\mathrm{p}}
\def\dd{\mathrm{d}}
\def\({\left(}
\def\){\right)}
\begin{document}

\begin{frontmatter}

\title{Coupling Temperature Distribution with the Single Particle Model}
\author[qdot]{Matthew Hunt}
\author[maths]{Florian Theil}
\author[maths,TFI]{Ferran Brosa Planella}
\author[wmg,TFI]{W. Dhammika Widanage\thanks{The authors acknowledge support from the EPSRC Prosperity Partnership (EP/R004927/1) and The Faraday Institution
(EP/S003053/1 Grant No. FIRG003)}}

\address[wmg]{WMG, University of Warwick, Coventry CV4 7AL}
\address[maths]{Mathematics Institute, Zeeman Building, University of Warwick, Coventry CV4 7AL}
\address[TFI]{The Faraday Institution, Quad One, Becquerel Avenue, Harwell Campus, Didcot, OX11 0RA}
\address[qdot]{Qdot Technology, Office F20 Atlas Building R27 Rutherford Appleton Laboratory Harwell, Didcot OX11 0QX}

\begin{abstract}
The DFN (Doyle-Fuller-Newman) model is well know for being accurate and computationally expensive. In situations where temperature gradients are important (eg fast charging) it is desirable to couple the temperature dynamics within a battery into the DFN model. This leads to even greater computational complexity. Inspired by the work of Marquis et al~ \cite{Marquis2019} we present the derivation of a reduced-order model based on the DFN model with temperature in the macroscale. The complexity of the reduced-order model is characterised by the local temperature plus one internal electro-chemical dimension and the electrolyte dynamics is accounted for by a simple correction term.\\[1em]
\end{abstract}


\end{frontmatter}

\section{Introduction}
With the rise of intermittent energy resources (wind, tidal and wave to mention a few) as well as the rise of electric and hybrid cars, there is a need to develop batteries with higher capacities, and the dominant technology is the lithium-ion battery. The modelling of lithium-ion batteries was initiated with the seminal paper by Doyle, Fuller and Newman \cite{Doyle1993, Fuller1994, Fuller1994a} which established the model known as the Doyle-Fuller-Newman (DFN) model. Although this model included temperature as a parameter, it ignored the dynamics of temperature based upon electrochemistry. When the battery is operated at a small C-rate, the temperature effects are relatively minor and the DFN model provides a good description of the charge/discharge process. At higher C-rates, the rise in temperature and its effects on the electrochemical properties are no longer negligible and the DFN model needs to be extended to account for them. Moreover, these changes in temperature enhance battery degradation \cite{Edge2021}, and these effects are particularly important in applications such as electric vehicles.

Electric vehicles are developing sophisticated thermal management systems. For example, the Toyota Prius relied on passive air cooling, while the latest Tesla Models rely on active liquid cooling design \cite{Yuksel2017}. Such thermal management systems rely on knowing the internal cell temperature distribution and its dynamics since they influence the design choices of the cooling arrangements (e.g. surface or tab cooling with cooling plates or use of immersed cooling etc.). Knowing the internal cell temperature dynamics also allows effective control algorithms to be developed to maintain a low temperature gradient ($\leq 5 ^\circ$C) both inside a cell and between the cells within a module \cite{Worwood2017}. Maintaining such low thermal gradients is crucial to ensure pack safety and minimise battery degradation. 

\begin{figure}[h!]
     \centering
     \begin{subfigure}[c]{0.47\textwidth}
         \centering
         \includegraphics[width=\textwidth]{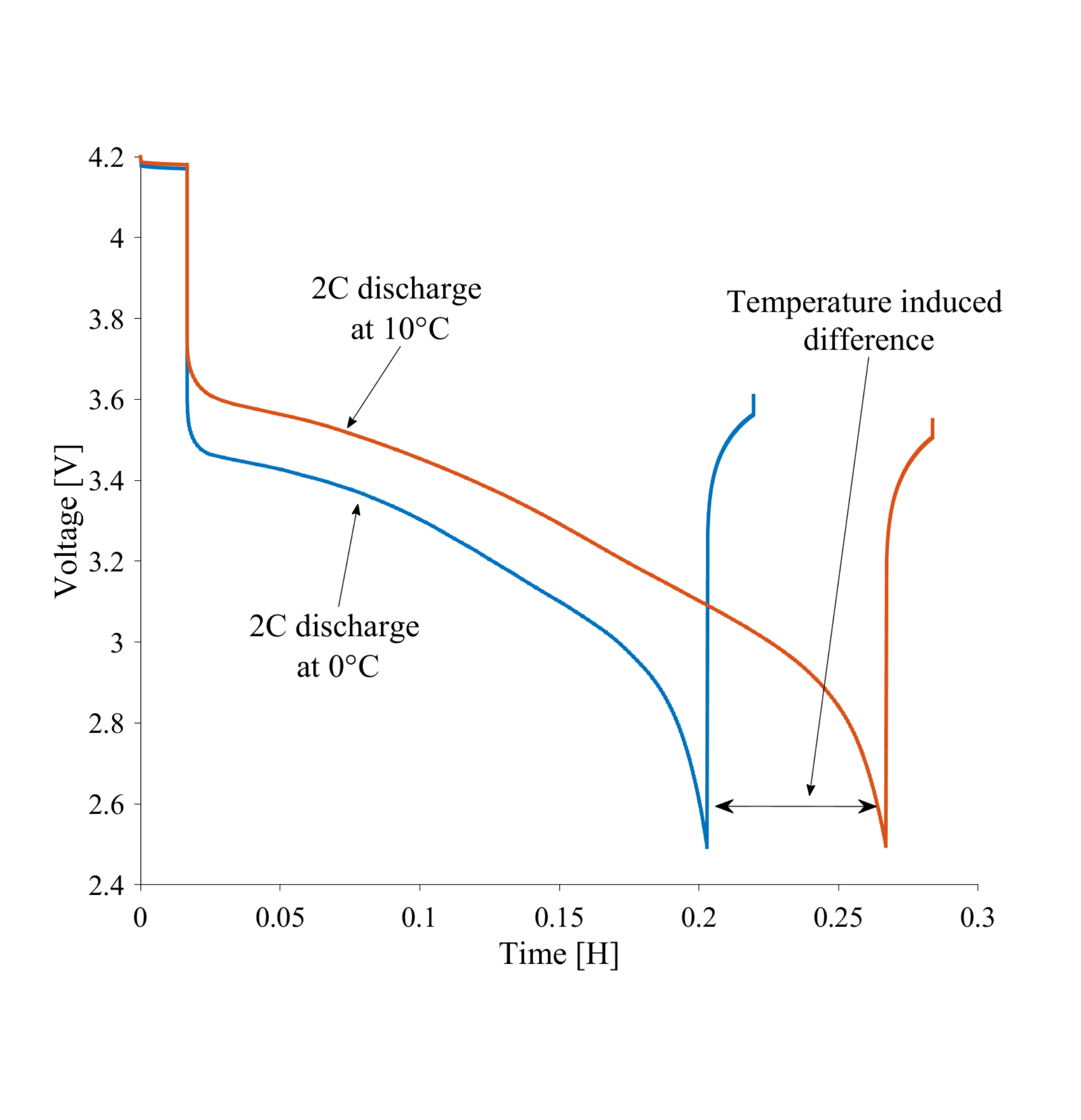}
         \caption{Influence of ambient temperature on a 2C discharge voltage response. The cells were thermally controlled at 0$^\circ$C and 10$^\circ$C in an oil bath.}
         \label{subfig:VolResp}
     \end{subfigure}
     \hfill
     \begin{subfigure}[c]{0.47\textwidth}
         \centering
         \includegraphics[width=\textwidth]{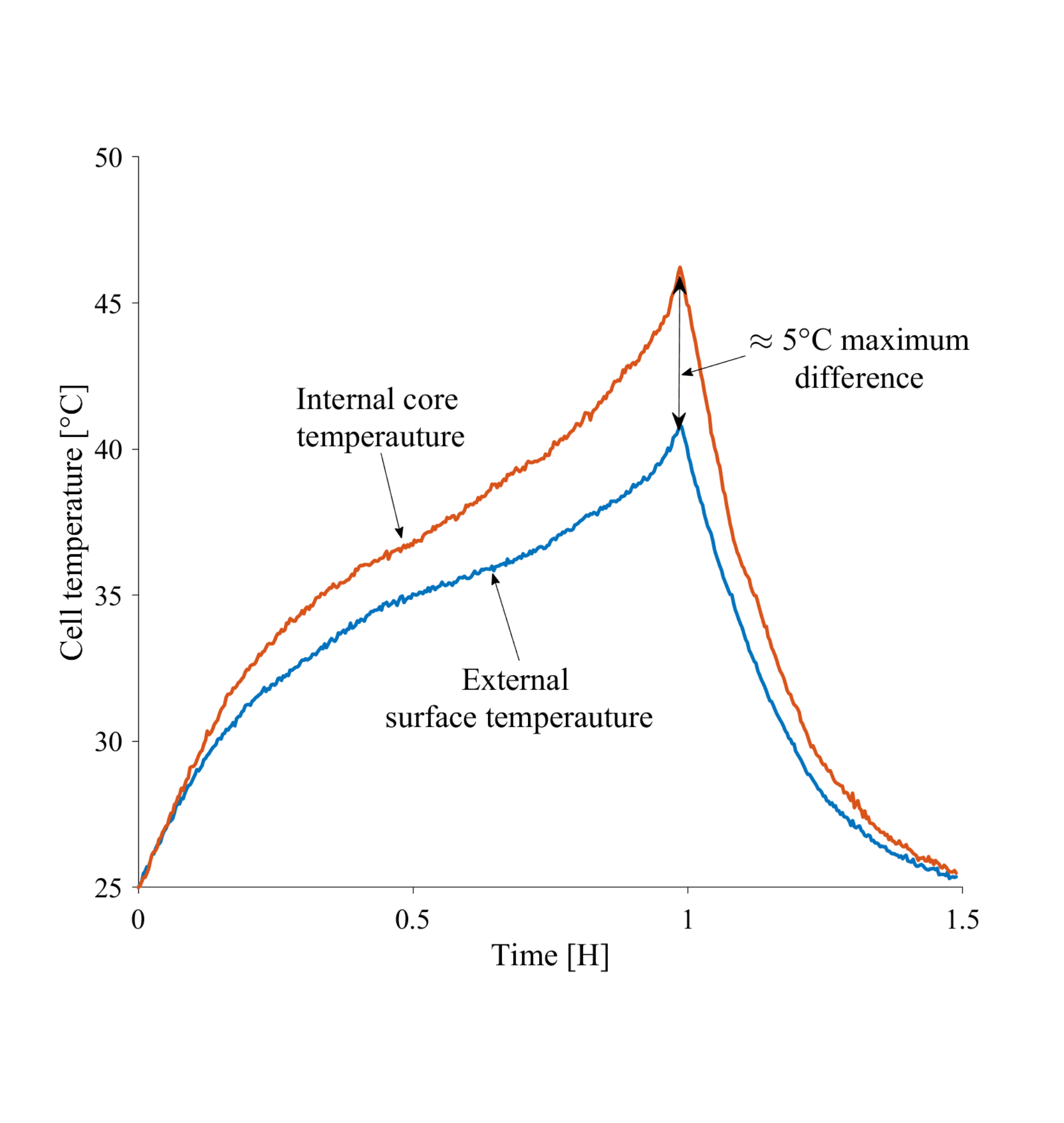}
         \caption{Temperature gradient build up in a cell when discharged at 1C at 25$^\circ$C ambient. The cell was thermally controlled in a thermal chamber with forced convection.}
         \label{subfig:TempResp}
     \end{subfigure}
        \caption{Voltage and thermal responses of a commercial 5 Ah 21700 cylindrical cell. The chemistry of the positive electrode is NMC811 while in the negative electrode it is graphite and silicon oxide. More details on this cell can be found in \cite{Chen2020}.}
        \label{fig:LGM50_Resp}
\end{figure}

Two measurement examples are provided below that show a typical gradient and the impact of ambient temperature on cell performance. Figure (\ref{subfig:VolResp}) demonstrates the increase in resistance when the ambient temperature of a commercial battery (5Ah NMC811 21700 cell \cite{Chen2020}) is cooled (0$^\circ$C compared to 10$^\circ$C). The cells on this occasion were thermally controlled in oil baths and a constant current discharge of 2C was applied. The voltage response while at 0$^\circ$C has a larger drop in potential and reaches the cut-off terminal voltage of 2.5V much earlier than when at 10$^\circ$C. This is due to the conductivity and diffusivity of the electrodes and electrolyte becoming less favourable when at lower temperatures (usually following an Arrhenius behaviour, \cite{Landesfeind2019,O'Regan2022}). Similarly Figure (\ref{subfig:TempResp}) demonstrates a typical temperature difference that occurs between the core of a cell (cylindrical) against the cell surface. The cell was placed in a thermal chamber with a set-point of 25$^\circ$ with forced convection and discharged at 1C. A maximum difference of approximately 5$^\circ$C is observed by the end of the discharge. The induced gradients will therefore lead to varied electrochemical kinetics (in the radial direction of the cell), which over a long period of use, will cause inhomogeneous ageing of the different layers within the cell.
\par\vspace{2mm}\noindent
\subsection{Our contribution}
The aim of this paper is to present an effective, physics-based model which tracks 
\begin{itemize}
\item The temperature distribution within the battery.
\item Lithium concentration in the electrodes.
\item Lithium concentration in the electrolyte.
\end{itemize}

The model can be viewed as an extension of the SPMe model by Marquis et al. \cite{Marquis2019} by including temperature as a dynamic variable. Here we provide a derivation via asymptotic expansion from 
the a standard DFN model with temperature, cf~\cite{Hunt2020}.

\subsection{Literature review}
The literature on temperature dynamics of a lithium ion battery is less developed than the account of electrochemical models. Depending on their level of accuracy, models either capture average temperatures or spatially resolved temperatures. There are three scales involved in battery modelling: the microscale, which is where the lithium ions diffuse in and out of the electrode particles; the mesoscale where the negative electrode, separator and positive electrode are considered (sometimes also called the \emph{cell level}); and the macroscale, which is the physical battery, a typically cylindrical, pouch cell or prismatic battery.
\begin{figure}
    \centering
    \includegraphics[scale=1]{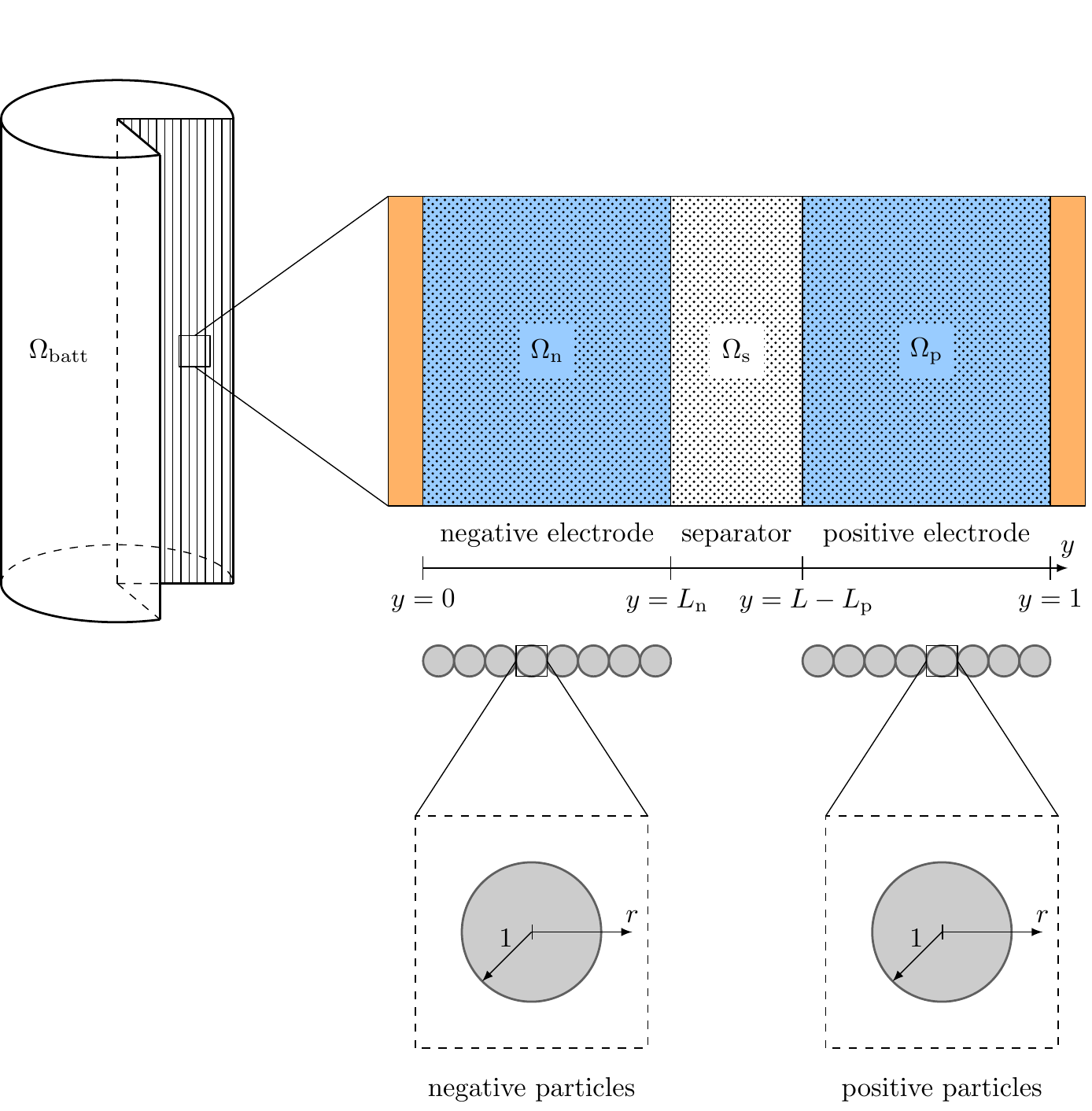}
    \caption{Illustration of the various scales and domains involved in the thermal DFN model (reproduced from \cite{BrosaPlanella2021}).}
    \label{scales}
\end{figure}


One of the first comprehensive approaches to write down a temperature equation for a lithium-ion battery was put forward in the seminal paper by Gu and Wang \cite{Gu2000} in 2000. The resulting model is for temperature on the mesoscale which is of the form of a general parabolic temperature equation with four source terms representing Ohmic heating in the solid phase, electrolyte, reversible and irreversible heating. The source terms are coupled to equations in the mesoscale in the form of the electric potentials.  

The paper \cite{Hennessy2020} presents a systematic derivation of a temperature for the macroscale and mesoscale. This is a comprehensive physics-based model which includes four heat source terms in the mesoscale and macroscale. An asymptotic reduction is carried based upon different timescales. 

There are a number of lumped parameter models, \cite{Lin2014} where the temperature is assumed to be homogeneous. The equation for temperature is an ordinary differential equation (ODE) rather than a partial differential equation (PDE), so temperature gradients within the battery on the macroscale are ignored. The models usually consider radiation and convection as mechanisms for cooling but for heating which is included in the ODE, the heating terms are usually a combination of two terms, Joule heating which is given by the usual $I^{2}_\mathrm{app}R$ term or a ``heat of polarisation'' term which is the voltage loss multiplied by the applied current. Sometimes reversible heating is included which may result in nonlinear governing equations.

There have been a number of approaches to include geometric effects, a notable example is the spiral geometry of the cell with cylindrical batteries, eg. \cite{Evans1989, Somasundaram2012, Guo2014, Psaltis2020}. These models make the assumption of an Archimedean spiral, is is generally noted that spiral geometry leads to an enhanced thermal conductivity. Quantitative estimates of this effect are reported in \cite{Psaltis2020}.

\subsection{A brief description of the model}
The model described in this article is a reduced order model of the more comprehensive model derived in \cite{Hunt2020} based on the analysis of the DFN model in \cite{Marquis2019}. One of main premises of the model is that the rate of diffusion in the electrolyte to the total discharge time is very small ($\sim 10^{-3}$) and this allows the reduction of the full thermal model and even removing one of the scales involved. Two other assumptions which are important is high conductivity in the electrode and the electrolyte. The resulting model has equations for the concentration of lithium in the electrode particles and the temperature distribution in the macroscale. It is noteworthy that the applied current plays the role of a time-dependent parameter in single-particle models like in~\cite{Marquis2019}. In our model the applied current is represented in the form of a time-dependent constraint. 

\section{Dimensionless Single Particle Model with Temperature distribution} \label{model_statement.sec}
The model accounts for the evolution of lithium concentration, electric potentials and temperature on two different scales with associated variables
\begin{description}
    \item[Macroscale] $x \in \Omega_\mathrm{batt}$ position in battery,
    \item[Microscale] $r\in [0,1]$ distance from centre of electrode particle.
\end{description}
The assumptions mentioned in the previous section allow for the analytical solution of all variables in the mesoscale. Only the remaining equations in the microscale and macroscale are required to be solved numerically. 

\subsection{Evolution of the battery temperature profile}
The temperature in the macroscale $T(x,t)$ is determined by the diffusion equation
\begin{equation} \label{eq1}
\partial_{t} T=\kappa \nab_x T +Q\quad \text{ in } \Ombatt,
\end{equation}
where $Q$ denotes the heat source and $\kappa$ is the thermal diffusivity. There are two geometries which are of interest, cylindrical and slab, and the use of the Laplacian covers both cases.
To model specific situations equation~\eqref{eq1} needs to be supplemented with suitable initial and boundary conditions.

The heat source $Q$ is the sum of of Ohmic heating in electrode particles and electrolyte $Q_\mathrm{Ohmic}$, and reversible and irreversible heating from the intercalation reaction ($Q_\mathrm{irr}$ and $Q_\mathrm{rev}$, respectively):
\begin{equation}
Q=Q_\mathrm{Ohmic}+Q_\mathrm{irr}+Q_\mathrm{rev}.
\end{equation}

Ohmic heating accounts for the electric resistance of the electrodes, and is given by
\begin{eqnarray*}
Q_\mathrm{Ohmic}& = &  \alpha\, i^{2},
\end{eqnarray*} 
where $i$ represents the local applied current density which has to be calculated and the value of $\alpha$ is determined in Section~\ref{ass_exp.sec} to be an exceptionally small contribution to the heat source, so we can effectively set $\alpha=0$. The formulae for reversible and irreversible heating are 
\begin{eqnarray*}
Q_\mathrm{\textrm{rev}}&=&(\Pi_\mathrm{n}-\Pi_\mathrm{p})\, i,\\
Q_\mathrm{\textrm{irr}}&=& \frac{2(1+\gamma T)\,i}{|\Omcell|}\left[\arsinh\left(\frac{i}{j_{0}\Ln}\right)+\arsinh\left(\frac{i}{j_{0}\Lp}\right)\right].
\end{eqnarray*} 
The function $j_{0}$ is called the exchange current density and defined by $j_{0}=\mu[(1-\cs)\cs\ce]^{\frac{1}{2}}$. The two closing conditions are
\begin{align}
\label{constvolt}
    \nabla_{x} V &= 0 \quad \text{ in } \Ombatt,\\
\label{charge_conservation}
\int_{\Omega_\mathrm{batt}} i(r,t) A(r)\,\dd r &= I_\mathrm{app}(t),
\end{align}
where $i(r)$ is the current density of the mesoscale cell with distance $r$ from the centre, $A(r)$ is the current collector area of the corresponding layer $A(r) = 2\pi h \lceil  \frac{r}{h} \rceil$ and $I_\mathrm{app}$ is the total current applied to the battery.

The meaning of \eqref{charge_conservation} is that the volume integral of the current density yields $I_\mathrm{app}$, the total applied current. From a physical perspective equation \eqref{charge_conservation} is the conservation of charge. The layers are treated as they are in parallel, this implies that the terminal voltage, $V$, across the cell will be the same for all layers. 

Thanks to~\eqref{constvolt} the value of terminal voltage only depends on time, ie $V=V(t)$. Within the framework of our first order approximation the terminals voltage is decomposed into a leading order value plus a correction, 
$$ V(t)=V^0(t)+\varepsilon \, V^1(t).$$
The value of $V^1$ is given by the formula~\eqref{defV1}, the leading order voltage $V^0$ and the current density $i$ are determined by via charge conservation  ~\eqref{charge_conservation} and the
requirement 
\begin{align}\label{V0eqn}
V^0 = \Phi_\mrs^0(y=1) - \Phi_\mrs^0(y=0),
\end{align}
with the convention that 
the values of the potential ${\Phis^0}$ are determined by $i$ via equations~\eqref{flux_cond} and the Butler-Volmer equation
\begin{align*}
\Phis^0 =& \frac{2}{\lambda}(1+\gamma T)\,\arsinh\left(\frac{J^0}{j_{0}} \right)+ \OCP(\cs|_{r=1}),
\end{align*}
with $\OCP$ being the open-circuit potential.
\subsection{Evolution of the lithium concentration in electrode particles}
The source terms for the temperature equation require the solution of the mesoscale, which can be done analytically thanks to the asymptotic assumptions. In \cite{Plett2015} the modelling in the electrolyte is 1D and this is the approach taken in this article. Then, twe have to solve numerically the microscale equations for the lithium concentration in the particles, which read
\begin{equation}
\p_{t}\cs=r^{-2}\p_{r}(r^{2}\Ds\p_{r}\cs) \quad r \in [0,1],
\end{equation}
with boundary and initial conditions:
\begin{equation} \label{flux_cond}
\Ds\p_{r}\cs=J^{0}\quad r=1,\quad \Ds\p_{r}\cs=0\quad r=0,\quad \cs(0,r)=1,
\end{equation}
where  
\begin{displaymath}
J^{0}=\left\{
\begin{array}{cc}
\frac{i}{\Ln}    & 0<y<\Ln, \\
0 &  \Ln<y<1-\Lp, \\
-\frac{i}{\Lp} & 1-\Lp<y<1,
\end{array}\right.
\end{displaymath}
and the parameters $\Ln$ and $\Lp$ represent the relative thickness of the negative and positive electrodes within the cell. This completes the statement of the model. 
\section{Derivation from DFN with Temperature} \label{derivation.sec}
In this section we demonstrate that the model can be obtained via asymptotic expansion from the standard DFN model with temperature (eg. \cite{Hunt2020} for a suitable choice of geometry).
The starting point is the dimensionless model derived in \cite{Hunt2020}. Note that all parameters except $\lambda$, $\gamma$, $\varepsilon$ are piecewise constant. The microscale model is the same as \cite{Gu2000}, \cite{Hennessy2020} and \cite{Evans1989}; the derivation mesoscale equations is based on asymptotic homogenisation, a method that aims to capture effects caused by the geometry of the microstructure. 

A slightly less involved approach which delivers similar equations is called
volume averaging. Here, the coefficients of the coarse grained equations are obtained in a statistical manner which is not necessarily fully consistent with an underlying microscopic model.

The main assumption is that the non-dimensional parameters $\De$, $\alphae$, $\alphas$ and $\nu$ are large and of roughly of the same size. To model these assumptions those parameters are multiplied by the common scaling factor $\frac{1}{\varepsilon}$. We will demonstrate that the effective model in Section~\ref{model_statement.sec} captures the limiting behaviour of the solutions up to first order when $\varepsilon$ is very small.

\subsection{Dimensionless DFN model}\label{nondimmod.sec}
With the above conventions the dimensionless DFN model with temperature reads as follows.
\subsubsection*{Microscale equations}
\begin{align}
\label{cseq}
\partial_t \cs &= \Ds r^{-2}\p_{r}\left(r^{2} \partial_r\cs\right)  && \mbox{ in } [0,1],\\
\label{csbc}
-\Ds \partial_r \cs &=\beta_\mrs J && \text{ if } r=1,
\end{align}
where $\Ds$ is the diffusion coefficient of lithium in the electrode (which may depend on the concentration itself,
space variables, and temperature), $J$ is the exchange current.
\subsubsection*{Mesoscale equations}
\begin{align}
 \label{ceeq}  \beta_\mre J=& \partial_t \ce-\frac{1}{\varepsilon}
    D_\mathrm{e} \nabla_y^2 \ce &&\mbox{ in } \Omega_\mathrm{cell},\\
\label{cebc}    0 =& \partial_y \ce && \text{ if } y \in \partial\Omcell,\\
\label{ceic} 1=&\ce &&\text{ if } t=0.
\end{align}
The exchange current $J$ and the overpotential $\eta$ are defined as
\begin{eqnarray}
J & = & j_{0}\sinh\left( \frac{\lambda}{2} \frac{\eta}{1+\gamma T}\right), \\
\eta & = &  \Phis-\Phie-\OCP(\cs|_{r=1}),
\end{eqnarray}
where $\mu$ is the reaction rate which is governed by the Arrhenius relation \cite{Plett2015}. 
The electric potentials $\Phis,\Phie$ are characterised by the equations
\begin{align} 
\label{phieeq} \varepsilon\,J  &= -  \alphae\,\partial_y^2\Phie +  \nu (1 + \gamma T)\,\partial_y^2 \log \ce&&  \Omcell,\\
\label{phiebc}0 &=\partial_y \Phie && \partial \Omcell,\\
\label{phiezero} 0&= \Phie(0),\\[0.5em]
\label{phiseq} \varepsilon\, J &=\alphas\,\partial_y^2\Phis && \Omcell\setminus\OmegaS, \\
\label{phisf} \varepsilon\, i &= -\alphas\, \partial_y \Phis && \partial \Omcell, \\
\label{phisnf}
0&=\partial_y \Phis && \partial \OmegaS,
\end{align}
where the parameters $\alphas$ and $\alphae$ may have different values in $\Omegap$, $\OmegaS$ and $\Omegan$.
\subsubsection*{Macroscale equations}
\begin{align*}
\partial_t T &= \nabla_x \cdot \left( \kappa \nabla_x T \right) + Q \quad \text{ in } \Ombatt,
\end{align*}
where
$$ Q = Q_\mathrm{Ohmic}+Q_\mathrm{irr}+Q_\mathrm{rev},$$
and
\begin{align}
   Q_\mathrm{Ohmic} =&  \frac{\lambda}{|\Omcell|} \int_{\Omcell} \left( \alpha_\mrs |\nabla_y \Phis|^2 + \alpha_\mre |\nabla_y \Phie |^2 - \nu (1 + \gamma T) \nabla_y \log \ce \cdot \nabla_y \Phie \right) \dd y,\\
    Q_\mathrm{rev} = & \frac{\lambda}{|\Omcell|} \int_{\Omcell} J \, \Pi \, \dd y,\\
    Q_\mathrm{irr} =& \frac{\lambda}{|\Omcell|} \int_{\Omcell} J \, \eta \, \dd y.
\end{align}

From \eqref{phiseq}-\eqref{phisnf} we can conclude
\begin{align}
    \label{charge_cons} i &=-\int_{\Omegan} J\, \dd y = \int_{\Omegap} J\, \dd y.
\end{align}
An important consequence of~\eqref{ceeq}, \eqref{cebc}, \eqref{ceic} and \eqref{charge_cons} is the equation
$$\int_{\Omcell}\ce\, \dd y =1.$$


\subsection{Asymptotic expansion}
\label{ass_exp.sec}
The goal is to expand the cell voltage
\begin{eqnarray*}
V= \Phis(1)-\Phis(0)= \Phis(t, x, y = 1) - \Phis(t, x, y = 0)
\end{eqnarray*}
into powers of $\varepsilon$, i.e.
$$ V = V^0 + \varepsilon\, V^1 +O(\varepsilon^2),\quad  0<\varepsilon\ll 1.$$

The main consequence of the high electrolyte conductivity assumption is that the mesoscale equations are eliminated. This holds not only at leading order, but also for the first order corrections.

We expand $\cs$, $\ce$ $\Phis$, $\Phie$ in powers of $\varepsilon$, i.e. $\ce = \ce^0 + \varepsilon \ce^1+O(\varepsilon^2)$ etc. Plugging this Ansatz into the non-dimensional model in Section~\ref{nondimmod.sec} yields the following equations.

\subsubsection*{Order $\varepsilon^0$}
\begin{eqnarray}
\ce^0 &=& 1,\\
\Phie^0 &=& 0\\
\label{defJ0}
J^0 &=& \begin{cases}
\frac{i(x)}{\Ln} & \Omegap,\\
0 & \OmegaS,\\
-\frac{i(x)}{\Lp} & \Omegan,
\end{cases}\\
\label{phi0}
\Phis^0 &=& \frac{2}{\lambda}(1+\gamma T)\,\arsinh\left(J^0\,\mu^{-1}[(1-\cs^0)\cs^0]^{-\frac{1}{2}} \right)+ \OCP(\cs^0),
\end{eqnarray}
and the PDE
\begin{align*}
\partial_t \cs^0 =& \Ds r^{-2} \partial_r (r^2 \partial_r \cs^0) & \mbox{ in } \OmegaS,\\
-\Ds \partial_r \cs^0 =& J^0 & \text{ on } \partial \OmegaS
\end{align*}
Note that both the initial and the boundary values of $\cs^0$ are constant in $\Omegan$ and $\Omegap$, therefore $\cs^0$ only depend on the microscale variable $r$, the macroscale variable $x$ but on the mesoscale variable $y$.

\subsubsection*{Order $\varepsilon^1$}
\begin{align} \label{ce1eqn}
 J^0&= D_\mathrm{e} \partial^2_y\ce^1, && 0=\partial_y \ce^1|_{\partial \Omcell}= \int_{\Omcell} \ce^1\, \dd y,\\
 \label{phie1eq}
J^0  &=\alphae\,\partial_y^2\Phie^1 &&  0=\partial_y \Phie^1|_{\partial \Omcell} = \Phie^1(0),\\[0.5em]
\label{phis1eq}
 J^0 &= \alpha_\mathrm{s} \,\partial_y^2 \Phis^1, && -i=\alphas\, \partial_y\Phis^1|_{\partial \Omcell}.
\end{align}
From~\eqref{charge_cons} we have that 
$\int_{\Omegap} J^1\, \dd y= \int_{\Omegan} J^1\, \dd y=0$, and by linearity of \eqref{cseq} and \eqref{csbc} one finds that
\begin{equation}
    \label{c1mean}
\int_{\Omegan}\cs^1\, \dd y =  \int_{\Omegap}\cs^1 \dd y =0.
\end{equation}
The solutions of eqns~\eqref{ce1eqn}, \eqref{phie1eq} and \eqref{phis1eq} are given by
\begin{align*}
\ce^1 &= \frac{i}{6\De} \left(\Lp^2 -\Ln^2\right)+\frac{i}{2\De} \begin{cases} \frac{y^2}{\Ln} - 1 + \Ln & \text{ if } y\in\Omegan,\\
2y-1 & \text{ if } y \in\OmegaS,\\
-\frac{1}{\Lp}(y-1)^2 +1-\Lp & \text{ if } y \in \Omegap,
\end{cases}\\
\Phie^1 &= \frac{i}{2\alphae} \begin{cases} \frac{y^2}{\Ln} & \text{ if } y\in\Omegan,\\
2y-\Ln & \text{ if } y \in\OmegaS,\\
-\frac{1}{\Lp}(y-1)^2 +2-\Ln-\Lp & \text{ if } y \in \Omegap,
\end{cases}\\
\Phis^1 &= \frac{i}{2\alphas} \begin{cases} \bPhisn^1+\frac{1}{\Ln}(y-\Ln)^2& \text{ if } y\in\Omegan,\\
\bPhisp^1-\frac{1}{\Lp}(y-1+\Lp)^2  & \text{ if } y \in \Omegap,
\end{cases}
\end{align*}
such that
\begin{align*}
\bcen^1 &= \frac{1}{|\Omegan|}\int_{\Omegan}\ce^1\,\dd y=\frac{i}{6\,\De}(\Lp^2 - \Ln^2 - 4\Ln + 3),\\
\bcep^1 &= \frac{1}{|\Omegap|}\int_{\Omegap}\ce^1\,\dd y=\frac{i}{6\,\De}(\Lp^2 - \Ln^2 + 4\Lp - 3),\\
\bPhien^1 &= \frac{1}{|\Omegan|}\int_{\Omegan} \Phie^1\, \dd y = \frac{\Ln}{6\alphae}\\
\bPhiep^1 &= \frac{1}{|\Omegap|}\int_{\Omegap} \Phie^1\, \dd y = \frac{1}{6\alphae}(6-4\Lp - 3\Ln)
\end{align*}
The values of $\bPhis^1$ on $\Omegap$ and $\Omegan$ are determined by~\eqref{charge_cons}:
\begin{align}\nonumber
  0 = &\frac{1}{|\Omegan|}\left(\int_{\Omegan}J\,\dd y - i\right)\\
  \nonumber
    =& \frac{\varepsilon}{2\,|\Omegan|}\int_{\Omegan} \biggl[\frac{\ce^1}{\ce^0}+ \frac{1-2\cs^0}{(1-\cs^0)\,\cs^0}\,\cs^1\\
    \nonumber
    &+\frac{\lambda}{1+\gamma T}\coth(\frac{\lambda}{2} \frac{\eta^0}{1+\gamma T})\left(\bPhis^1-\bPhie^1+\OCP'(\cs^0)\,\cs^1\right)\biggr]\,J^0\, \dd y +O(\varepsilon^2)\\
    =& \frac{\varepsilon }{2}\,\left[ \bcep^1 +\frac{\lambda}{1+\gamma T}\coth(\frac{\lambda}{2} \frac{\eta^0}{1+\gamma T})\left(\bPhis^1-\bPhie^1
    \right)\right] ,
\end{align}
and analogous for $\Omegan$. The terms involving $\cs^1$ drop out thanks to~\eqref{c1mean}. This implies that
\begin{align*}
 \bPhis^1& = \bPhie^1 - \frac{1}{\lambda}(1+\gamma T)\tanh\left(\frac{\lambda}{2} \frac{\eta^0}{1+\gamma T}\right) \bce^1\\
 &= \bPhie^1 - \frac{1}{\lambda}(1+\gamma T)\left(1+\sinh^{-2}\left(\frac{\lambda}{2} \frac{\eta^0}{1+\gamma T}\right)\right)^{-\frac{1}{2}}\bce^1\\
 &= \bPhie^1 - \frac{1}{\lambda}(1+\gamma T)\left(1+\frac{1}{i^2} (1-\cs^0)\cs^0\right)^{-\frac{1}{2}}\bce^1.
\end{align*}
Putting the results of these calculations together we obtain the
correction term
\begin{align}\nonumber 
    V^1 =& \Phis^1(1)-\Phis^1(0)= 
    \frac{i}{2 \alphas}\left(\bPhisp^1-\bPhisn^1-\Lp-\Ln\right) \\\nonumber
    =& \frac{i}{2 \alphas}\Biggl[\frac{1}{\alphae}\left(1-\frac{2}{3}(\Lp-\Ln)\right)-\Lp-\Ln\\\nonumber
    & +\frac{1}{\lambda}(1+\gamma T)\Biggl(\left(1+\frac{1}{i^2} (1-{\cs^0}_\mathrm{,n})\,{\cs^0}_\mathrm{,n}\right)^{-\frac{1}{2}} \,\frac{i}{6\,\De}(\Lp^2 - \Ln^2 - 4\Ln + 3)\\
    &- \left(1+\frac{1}{i^2} (1-{\cs^0}_\mathrm{,p})\,{\cs^0}_\mathrm{,p}\right)^{-\frac{1}{2}}\,\frac{i}{6\,\De}(\Lp^2 - \Ln^2 + 4\Lp - 3)\Biggr)\Biggr].
    \label{defV1}
\end{align}
Note that $V^1$ does not depend on $x$, but $i$ and $\cs^0$ are functions of $x$. To evaluate \eqref{defV1} one can choose an arbitrary point $x \in \Ombatt$.

It is now possible to compute the heat source for the macroscale. The heat source will be different depending on the region of the cell it is in. The heat source in each region is given by

\begin{eqnarray*}
    	Q_\mathrm{s} & = & \frac{\lambda\alphas}{|\Omcell|}\int_{\Omcell}|\p_{y}\Phis|^{2} \dd y \\
	& = & \frac{\lambda\alphas}{|\Omega_\mathrm{cell}|}\int_{\Omcell}(0+\varepsilon\p_{y}\Phis^{1})^{2} \dd y \\
	& = & \frac{\lambda\alphas\varepsilon^{2}}{|\Omcell|}\int_{\Omcell}(\p_{y}\Phis^{1})^{2} \dd y.
\end{eqnarray*}
A similar calculation can be carried out for $Q_\mathrm{e}$ and so the heat source term due to Ohmic heating is negligible as compared to the reversible and irreversible heating. In \cite{Lin2014}, much of the input from the electrochemical part of the model is the ``$I_\textrm{app}^{2}R$'', as shown here, this is a relative small effect compared to other effects. It may be expected that models which essentially use just Ohmic heating as a source term will likely give poor predictions.  
The only function of $y$, the mesoscale variable is $c_{\textrm{e}}$ which appears in the function $j_{0}=K\sqrt{c_{\textrm{s}}-c_{\textrm{s}}^{2}}\sqrt{1+\varepsilon c_{\textrm{e}}^{1}}$, where $K$ is the reaction rate, which is dependent upon the temperature. So we expand the mesoscale heat source term to $O(1)$ and integrate over the cell getting:
\begin{equation} \label{defalpha2}
Q_\mathrm{irr}=\frac{2(1+\gamma T)}{|\Omcell|}\left[\Ln\arsinh\left(\frac{i}{j_{0}\Ln}\right)+\Lp\arsinh\left(\frac{i}{j_{0}\Lp}\right)\right],
\end{equation}

\section{Conclusions}

In this work, we have provided a coupled thermal-electrochemical model which captures the temperature distribution across the battery. The model has been formally derived from the fully coupled thermal Doyle-Fuller-Newman model which involves three very distinct scales: battery, cell and particle. Using asymptotic methods we reduced the full model taking the assumptions of fast diffusion of ions in the electrolyte, and high conductivity both in the electrodes and the electrolyte. These assumptions allow us to solve analytically at the mesoscale (cell level) leaving only the temperature at the macroscale (battery level) and lithium concentration at the microscale (particle level) to be solved numerically. The resulting model, which belongs to the family of Single Particle Models, captures the behaviour of the battery at three different scales with a similar computational cost to the isothermal Doyle-Fuller-Newman model. This simple model, even though it is still too complex to be implemented in battery management systems, can play a very important role in demanding applications such as battery design and diagnosis.

\section{Acknowledgements}
The authors acknowledge support from the EPSRC Prosperity Partnership (EP/R004927/1) and The Faraday Institution (EP/S003053/1 Grant No. FIRG025).

\appendix
\section{Dimensional DFN model}\label{dimmod.sec}
We state the dimensional version of the DFN model with temperature which has been derived in~\cite{Hunt2020}. Dimensional variables and parameters are consistently written with a `hat' (e.g. $\hat{D}_\mre$) to prevent confusion with dimensionless versions.
\subsubsection*{Microscale equations}
\begin{eqnarray}
\label{cseq_dim}
\partial_{\hat t} \hat c_\mrs &=& \hat \Ds \nabla_{\hat z}^2 \hat c_\mrs \quad \mbox{ in } \hat \Omega_\mrs,\\
\label{csbc_dim}
-\hat \Ds \nabla_{\hat z} \hat c_\mrs\cdot {\nu_\mrs} &=& \frac{\hat J}{a F} \quad \mbox{ on } \partial \hat \Omega_\mrs,
\end{eqnarray}
where $\hat \Ds$ is the diffusion coefficient of lithium in the electrode (which may depend on the concentration itself, space variables, and temperature), $\hat J$ is the interfacial current, $a$ is the surface area density and $F$ is the Faraday constant.

\subsubsection*{Mesoscale equations}
\begin{eqnarray}
 \label{ceeq_dim}  \frac{\hat J}{F}&=& \varphi_\mathrm{e}\,\partial_{\hat t} \hat c_\mre- 
    \hat D_\mathrm{e}\nabla_{\hat y}^2 \hat c_\mre \mbox{ in } \hat \Omega_\mathrm{cell}\\
\label{cebc_dim}    0 &=&n \cdot \nabla_{\hat y} \hat c_\mre|_{\hat y \in \{0,L\}},\\
\label{ceic_dim} c_{\mre,\mathrm{init}} &=&\hat c_\mre|_{\hat t=0}.
\end{eqnarray}
The exchange current $\hat J$ and the overpotential $\hat \eta$ are defined as
\begin{eqnarray}
\hat J &=& \begin{cases} a \, m \left[\left(c_\mrs^{\max}-\hat c_\mrs\right)\, \hat c_\mrs \, \hat c_\mre\right]^\frac{1}{2}\sinh\left( \frac{F}{2 R_\mathrm{g} T} \hat \eta\right), & \hat \Omega_\mathrm{cell} \setminus \hat\Omega_{\mathrm{S}},\\
0 & \hat \Omega_\mathrm{S}
\end{cases}
\\
\hat \eta &=&  \hat \Phi_\mrs- \hat \Phi_\mre- \hat U_\mathrm{eq}(\hat c_\mrs).
\end{eqnarray}
The electric potentials $\hat \Phi_\mrs,\hat \Phi_\mre$ are characterised by the equations
\begin{align} 
\label{phieeq_dim} \hat J  &= -\sigma_\mre \nabla_{\hat y}^2\hat \Phi_\mre +  2 (1 - t^+) \sigma_\mre \frac{R_\mathrm{g} T}{F} \,\nabla_{\hat y}^2 \log \hat c_\mre&&  \hat\Omega_\mathrm{cell},\\
\label{phiebc_dim}0 &=n\cdot\nabla_{\hat y} \hat \Phi_\mre && \partial \hat \Omega_\mathrm{cell},\\
\label{phiezero_dim} 0&= \hat \Phi_\mre(0),\\[0.5em]
\label{phiseq_dim} \hat J &= \sigma_\mrs \nabla_{\hat y}^2\hat \Phi_\mrs && \hat \Omega_\mathrm{cell}\setminus \hat \Omega_\mathrm{S} \\
\label{phisf_dim} \mp \varepsilon\, \hat I &= \sigma_\mrs n\cdot \nabla_{\hat y} \hat \Phi_\mrs && \partial \hat \Omega_\mathrm{cell}, \\
\label{phisnf_dim}
0&=n\cdot \nabla_{\hat y} \hat \Phi_\mrs && \partial \hat \Omega_\mathrm{S},
\end{align}
where the parameters $\sigma_\mrs$ and $\sigma_\mre$ are may have different values in $\hat \Omega_\mathrm{p}$, $\hat \Omega_\mathrm{S}$ and $\hat \Omega_\mathrm{n}$. 

\subsubsection*{Macroscale equations}
\begin{align*}
\theta \partial_{\hat t} \hat T &= \nabla_{\hat x} \cdot \left( \hat \kappa \nabla_{\hat x} \hat T \right) + Q,
\end{align*}
where
$$ Q = Q_\mathrm{Ohmic}+Q_\mathrm{irr}+Q_\mathrm{rev}$$
and
\begin{align*}
   Q_\mathrm{Ohmic} =&  \frac{1}{\| \hat \Omega_\mathrm{cell} \|}\int_{\hat \Omega_\mathrm{cell}} \left( \sigma_\mrs \| \nabla_{\hat y} \hat \Phi_\mrs \|^2  + \sigma_\mre \| \nabla_{\hat y} \hat \Phi_\mre \|^2 - 2 (1 - t^+) \sigma_\mre \frac{R_\mathrm{g} T}{F} \nabla_{\hat y} \log \hat c_\mre \cdot \nabla_{\hat y} \hat \Phi_\mre \right) \dd x, \\
   Q_\mathrm{rev} = & \frac{1}{\| \hat \Omega_\mathrm{cell} \|}\int_{\hat \Omega_\mathrm{cell}} \hat J \hat \Pi \dd x,\\
   Q_\mathrm{irr} =& \frac{1}{\| \hat \Omega_\mathrm{cell} \|}\int_{\hat \Omega_\mathrm{cell}} \hat J \hat \eta \dd x.
\end{align*}

We define the following scalings of the problem
\begin{equation}\label{eq:scalings}
\begin{aligned}
\hat t &= t_0 t, & \hat c_\mrs &= c^{\max} c_\mrs, & \hat \Phi_\mrs &= \Phi_0 \Phi_\mrs, & \hat J &= \frac{I_0}{L} J, \\ 
\hat z &= R z, & \hat c_\mre &= c_{\mre,\mathrm{init}} c_\mre, & \hat \Phi_{\mre} &= \Phi_0 \Phi_{\mre}, & \hat I &= I_0 I,\\
\hat y &= L y, & \hat \eta &= \Phi_0 \eta, &  \hat U_k &= \Phi_0 U_k, & \hat T &= \frac{R_\mathrm{g} T_\mathrm{amb} c_{\mathrm{n}}^{\max}}{\theta} T + T_\mathrm{amb}, \\
\hat x &= L_\mathrm{batt} x, & \Omega_\mathrm{batt} &= L_\mathrm{batt} \hat \Omega_\mathrm{batt}, & \Pi &= \frac{R_\mathrm{g} T_\mathrm{amb} }{F} \hat \Pi, & \hat Q &= \frac{i_0 R_\mathrm{g} T_\mathrm{amb}}{L F} Q,
\end{aligned}
\end{equation}
where $t_0 = F c_\mathrm{n}^{\max} L / I_0$. Note that some of the dimensional parameters here, such as $R$, $\varphi_\mre$, $\sigma_\mrs$ or $c^{\max}$, are piecewise constant to capture the different properties of each electrode and the separator.

The parameters of the dimensionless DFN model in section~\ref{model_statement.sec} are given by
\begin{equation}\label{eq:ND_groupings}
\begin{aligned}
\alpha_\mrs &= \frac{\sigma_\mrs \Phi_0}{L I_0}, & \alpha_\mre &= \frac{\sigma_\mre \Phi_0}{L I_0}, & \beta_\mrs &= \frac{I_0 t_0}{a R F c^{\max} L}, & \beta_\mre &= \frac{I_0 t_0 (1 - t^+)}{\varphi_\mre F c_{\mre, \mathrm{init}} L},\\
\Ds &= \frac{\hat \Ds t_0}{R^2}, & D_\mre &= \frac{\hat D_\mre t_0}{\varphi_\mre L^2}, & \gamma &= \frac{\Delta T}{T_\mathrm{amb}}, & \mu &= \frac{a \, m \, c^{\max} \sqrt{c_{\mre, \mathrm{init}}} L}{I_0},\\
\lambda &= \frac{\Phi_0 F}{R_\mathrm{g} T_\mathrm{amb}}, & \kappa &= \frac{\hat \kappa t_0}{\theta L_\mathrm{batt}^2} &  \nu &= 2 (1 - t^+) \frac{\sigma_\mre R_\mathrm{g} T_\mathrm{amb}}{F L I_0}. 
\end{aligned}
\end{equation}
\newpage
\bibliographystyle{model1-num-names}
\bibliography{references}
\end{document}